\documentclass{statsoc}
\usepackage[a4paper]{geometry}
\usepackage{graphicx}
\pdfoutput=1
\usepackage[textwidth=8em,textsize=small]{todonotes}
\usepackage{amsmath}
\usepackage{natbib}
\usepackage{amsfonts}
\usepackage{booktabs}
\usepackage{multirow}
\usepackage{rotating,tabularx}
\usepackage{caption}

\title[Incorporating increased variability in cancer biomarker discovery]{Incorporating increased variability in testing for cancer DNA methylation}
\author[James Y. Dai {\it et al.}]{\small James Y. Dai$^{1,2}$, Heng Chen$^{3}$, Xiaoyu Wang$^{1}$,  Wei Sun$^{1,2}$, Ying Huang$^{1,2}$ , William M. Grady$^{1,4}$, Ziding Feng$^{1,2}$}
\vspace{15pt}
\address{\small
Public Health Sciences Division, Fred Hutchinson Cancer Research Center, Seattle, U.S.A.$^{1}$\\
Department of Biostatistics, University of Washington, Seattle, U.S.A.$^{2}$\\
Department of Biostatistics, Gilead Sciences, Seattle, U.S.A.$^{3}$\\
Clinical Research Division, Fred Hutchinson Cancer Research Center, Seattle, U.S.A.$^{4}$
}
\vspace{15pt}
\email{\small jdai@fredhutch.org}

\begin{document}
\begin{abstract}
Cancer development is associated with aberrant DNA methylation, including increased stochastic variability. Statistical tests for discovering cancer methylation biomarkers have focused on changes in mean methylation. To improve the power of  detection, we propose to incorporate increased variability in testing for cancer differential methylation by two joint constrained tests: one for differential mean and increased variance, the other for increased mean and increased variance. To improve small sample properties, likelihood ratio statistics are developed, accounting for the variability in estimating the sample medians in the Levene test. Efficient algorithms were developed and  implemented in \texttt{DMVC} function of {\bf R} package \texttt{DMtest}. The proposed joint constrained tests were compared to standard tests and partial area under the curve (pAUC) for the receiver operating characteristic curve (ROC) in simulated datasets under diverse models. Application to the high-throughput methylome data in The Cancer Genome Atlas (TCGA) shows substantially increased yield of candidate CpG markers.
\end{abstract}

\section{Introduction}
\label{sec1}

Cancer development is associated with profound modifications in the epigenome, a multi-layer regulatory infrastructure for gene expression and cellular lineage \citep{Esteller2007,Esteller2008,Baylin2011,Shen2013a}. The most studied cancer epigenetic alteration to date is the 5-cytosine methylation at CpG dinucleotides. Occurring early in carcinogenesis and biochemically more stable than RNA transcripts, aberrant DNA methylation has become an important molecular target for developing cancer early detection marker \citep{Laird2003,Issa2008}.



High-throughput assays such as the Illumina Infinium HumanMethylation450, EPIC BeadChips and whole genome bisulfite sequencing (WGBS) enable interrogation of cancer methylome for CpG biomarker candidates \citep{Hao2017}. The current statistical engine for detecting cancer aberrant methylation entails testing equal means between cancer and normal samples for a single CpG site (differentially methylated CpG, {\bf DMC}) or a region with multiple adjacent CpGs (differentially methylated region, {\bf DMR}), as implemented in popular software such as \texttt{Minfi} \citep{Aryee2014} and \texttt{ChAMP} \citep{Morris2014}. The high dimensionality of genome-wide CpGs ($>$500,000) and typically small sample size for biomarker discovery studies limit the statistical power to identify novel markers.

\begin{figure}[!ht]
  \vspace{15pt}
\centering
  \includegraphics[width=.75\textwidth]{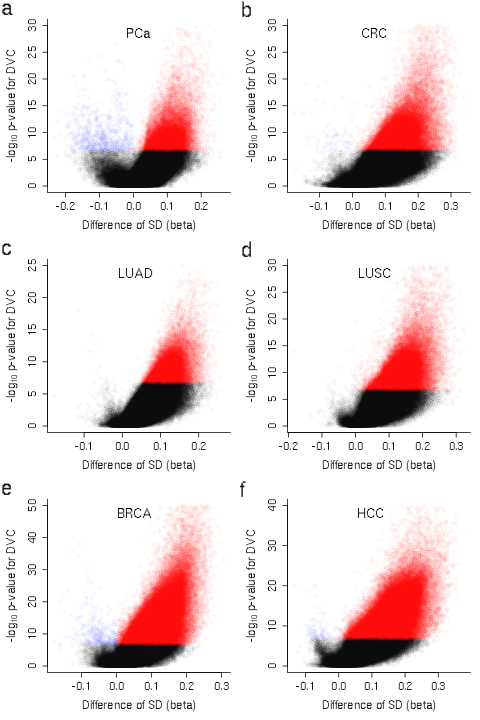}
        \caption{Volcano plots for testing DVC between cancers and normal samples in six major cancers from TCGA. Red circles and blue circles are hypervariable and hypovariable DVCs with family-wise error rate $<$ 0.05.  (a) Prostate cancer. (b) Colorectal cancer. (c) Lung adenocarcinoma. (d) Lung squamous cell carcinoma. (e) Breast cancer. (f) Liver hepatocellular carcinoma. }
\label{volcano_plot_six_cancers}
      \end{figure}

Cancer is a heterogeneous disease. Differential cancer methylation is also evident by increased stochastic variability. This has been observed across cancers \citep{Hansen2011,Phipson2014}, which may reflect adaptation to local tumor environments in the carcinogenesis process. Testing for equal variances in sample groups has been studied in the statistical literature for decades \citep{Brown1974}. Levene test achieves a balance between sensitivity and robustness to outliers, therefore has been used for detecting differentially variable CpG ({\bf DVC}) \citep{Phipson2014}.  Figure \ref{volcano_plot_six_cancers}  shows the volcano plots for testing differential variability in six major cancers available in TCGA using Levene test: namely prostate cancer (PCa),  colorectal cancer (CRC), lung squamous cell carcinoma (LUSC), and lung adenocarcinoma (LUAD),  breast cancer (BRCA), and liver hepatocellular carcinoma (HCC). A striking observation is that nearly all DVCs show increased variability in cancer samples ($>$99\% in BRCA, CRC, HCC, LUSC and LUAD, $>$95\% in PCa), confirming that cancer DVC is ubiquitously hypervariable. More results will be presented in Table 2. 

In this work, we develop joint tests that combine differential means and increased variances in CpG methylation data, aiming to improve the power of detecting cancer biomarker candidates. Specifically gearing to the ubiquitously increased variability in cancer DNA methylation, we develop two constrained hypothesis tests: a test for {\bf differential} mean and {\bf increased} variance, and another test for {\bf increased} mean and {\bf increased} variance. We used the estimating equation theory and likelihood ratio tests for constrained hypothesis to construct test statistics. We found that the variability in estimating the medians in Levene test needs to be accounted for in test statistics, in order to achieve better control of the type I error rate for high dimensional p-values. Computationally efficient algorithms for these tests were developed in the \texttt{DMVC} function of {\bf R} package \texttt{DMtest}, which is capable of scanning 500,000 CpGs for biomarker leads in a few minutes.


Another goal of this work is to compare the proposed constrained joint tests to the partial area under the curve (pAUC) measure for receiver operating characteristics curve (ROC), that is commonly used in the biomarker literature for selecting cancer gene expression markers \citep{Pepe2003}. In a similar one-sided fashion, pAUC evaluates the discriminative performance between cancers and controls in the high specificity area of the receiver operating characteristic (ROC) curve. In simulated datasets, we compare error control and power performance of pAUC, the standard t-test, Levene test, and the proposed constrained tests in diverse models. The utility of the proposed constrained tests for biomarker discovery will be investigated by TCGA methylome data.

\section{Methods}
\label{sec2}

\subsection{Joint test of differential mean and differential variance in CpG methylation}

 To develop the constrained hypothesis testing, it is necessary to first consider the standard null hypothesis for testing equal means and equal variances. Suppose there are $n$ samples with DNA methylome data available for the tumor-normal comparison, denoted by $(\mbox{\bf Y}_i,X_i,W_i)$, for $i=1,...,n$, where $\mbox{\bf Y}_i$ is a length $p$ vector for methylation M-values in $p$ CpG sites, $X_i$ is the indicator for cancer ($X_i=1$) or normal sample ($X_i=0$), and $W_i$ is the vector of additional covariates such as age and gender that should be adjusted for in the following regression analysis. Let $Y_{ij}$ denote M-value of the $j^{th}$ CpG site for the $i^{th}$ sample . DMC and DVC are tested separately by fitting linear regression models:
\begin{eqnarray}
  \mathbb{E}(Y_{ij})&=& \beta_{0j} + \beta_{1j} X_i + \beta_{2j} W_i,  \label{eq:model1} \\
  \mathbb{E}(|Y_{ij} - \widetilde{m}_{g(i)j}|)&=& \alpha_{0j} + \alpha_{1j} X_i + \alpha_{2j}W_i,  \label{eq:model2}
\end{eqnarray}
where $g(i)$ is the group (tumor or normal) label for $i^{th}$ sample, $\widetilde{m}_{g(i)j}$ is the sample median for $j^{th}$ CpG site in $g(i)$ group, $|Y_{ij} - \widetilde{m}_{g(i)j}|$ is the absolute difference to the corresponding group median.  Note that model (\ref{eq:model2}) implements the Levene test for homogeneous variances in two groups \citep{Brown1974}. The standard null hypothesis for testing equal means and equal variances is
\[
\mbox{H}_{0}: \beta_{1j}=0, \alpha_{1j}=0, \mbox{\hspace{10pt} versus \hspace{10pt}}\mbox{H}_{1a}: \beta_{1j} \neq 0 , \alpha_{1j} \neq 0,
\]
which often entails a 2-df Wald test.

In general, the outcomes of the above regression models $(Y_{ij},|Y_{ij} - \widetilde{m}_{g(i)j}|)$ are correlated, except in one special scenario where the distribution of $Y_{ij}$ is symmetric. For simplicity, we showcase the special case between $Y_{ij}$ and $|Y_{ij} - m_{g(i)j}|$ assuming $m_{g(i)j}$ is the population median for $j^{th}$ CpG site in $g(i)$ group.
\begin{eqnarray}
 & & \text{cov}\left(Y_{ij},|Y_{ij} - m_{g(i)j}|\right) \nonumber \\
 &=& \mathbb{E}\left(Y_{ij}|Y_{ij} - m_{g(i)j}|\right) - \mathbb{E}\left(Y_{ij}\right)\mathbb{E}\left(|Y_{ij} - m_{g(i)j}|\right) \nonumber \\
 &=&\mathbb{E}\left[(Y_{ij} - m_{g(i)j})|Y_{ij} - m_{g(i)j}|\right]
    + \mathbb{E}\left[(m_{g(i)j} - \mathbb{E}(Y_{ij}))|Y_{ij} - m_{g(i)j}|\right] \nonumber \\
&=&\frac{1}{2}\mathbb{E}\left[(Y_{ij} - m_{g(i)j})^2|Y_{ij} - m_{g(i)j}>0\right] - \frac{1}{2}\mathbb{E}\left[(Y_{ij} - m_{g(i)j})^2|Y_{ij} - m_{g(i)j}<0\right] \nonumber \\
& & + \mathbb{E}\left[(m_{g(i)j} - \mathbb{E}(Y_{ij}))|Y_{ij} - m_{g(i)j}|\right] = 0 \nonumber
\end{eqnarray}
where the first two terms are cancelled out because of symmetry of $Y_{ij}$ and the third term becomes zero because $m_{g(i)j} = \mathbb{E}(Y_{ij})$ under null hypothesis with the symmetric property. Other than the special case, the 2-df Wald test for testing $\mbox{H}_{0}: \beta_{1j}=0, \alpha_{1j}=0$ needs to account for the correlation. The asymptotic bivariate Gaussian distribution of $(\hat{\beta}_{1j}, \hat{\alpha}_{1j})$ can be derived from the estimating equation theory, and the asymptotic variance matrix $\Sigma_j$ for ($\hat{\beta}_{1j}$, $\hat{\alpha}_{1j}$) can be computed by the robust sandwich estimator. Specifically, let $u_{i1}$ be the estimating function for the first regression model (\ref{eq:model1}) and $u_{i2}$ be the estimating function for the second regression model (\ref{eq:model2}). Let $\mbox{\boldmath $ \beta $}_{j} =(\beta_{0j},\beta_{1j},\beta_{2j}) $ and $\mbox{\boldmath $ \alpha $}_{j} =(\alpha_{0j},\alpha_{1j},\alpha_{2j})$. When sample size is sufficiently large,  ($\widehat{\mbox{\boldmath $ \beta $}}_{j}$, $\widehat{\mbox{\boldmath $ \alpha $}}_{j}$) follows a Gaussian distribution with mean ($\mbox{\boldmath $ \beta $}_{j}$  ,$\mbox{\boldmath $ \alpha $}_{j}$) and variance matrix $\Sigma_j$, expressed as followed,
\begin{eqnarray}
\sqrt{n} \left(\begin{array}{c} \widehat{\mbox{\boldmath $ \beta $}}_{j}  - \mbox{\boldmath $ \beta $}_{j} \\ \widehat{\mbox{\boldmath $ \alpha $}}_{j} - \mbox{\boldmath $ \alpha $}_{j} \end{array} \right) \rightarrow_d \mathcal{N}\left(0, \left[ \begin{array}{cc} I(\mbox{\boldmath $ \beta $}_{j})^{-1} \mathbb{E} (u_{i1}^Tu_{i1}) I(\mbox{\boldmath $ \beta $}_{j})^{-1} & I(\mbox{\boldmath $ \beta $}_{j})^{-1} \mathbb{E} (u_{i1}^Tu_{i2}) I(\mbox{\boldmath $ \alpha $}_{j})^{-1} \\ I(\mbox{\boldmath $ \beta $}_{j})^{-1} \mathbb{E} (u_{i2}^Tu_{i1}) I(\mbox{\boldmath $ \alpha $}_{j})^{-1} & I(\mbox{\boldmath $ \alpha $}_{j})^{-1} \mathbb{E} (u_{i2}^Tu_{i2}) I(\mbox{\boldmath $ \alpha $}_{j})^{-1} \end{array} \right] \right),  \label{asymp}
\end{eqnarray}
where $I(\mbox{\boldmath $ \beta $}_{j}) =  \mathbb{E} (\partial u_{i1}/\partial \mbox{\boldmath $ \beta $}_{j})$ and $I(\mbox{\boldmath $ \alpha $}_{j}) =  \mathbb{E} (\partial u_{i2}/\partial \mbox{\boldmath $ \alpha $}_{j})$.

A cautionary note is that the derivation treats the sample medians in the two groups $\widetilde{m}_{g(i)j}$ as if they are known quantities. When sample size is small, however, we notice that the Wald test using the asymptotic covariance matrix ignoring the variability associated with sample medians appears to cause inflation of the type I error, particularly for the small p-values needed for high-dimensional testing. Table 1 shows the empirical type I error and the nominal p-value cutoff for p-value cut-off 0.05 or 0.001 in simulated datasets from 4 different models. At the 0.05 level, the standard 2-df test already shows inflated type I error rates (numbers in red). The inflation is worsened at the 0.001 level, reaching as high as 5-fold inflation for some models.

The estimating equation derivation can account for the extra variability due to estimation of sample medians. Observe that the estimating function for $\mbox{\boldmath $ \alpha $}_{j}$ can be decomposed to two components,
\begin{eqnarray}
 & & \frac{1}{\sqrt{n}}\sum_i u_{2i} \nonumber \\
 &=& \frac{1}{\sqrt{n}} \left\{ \sum_{i: Y_{ij}\geq \widetilde{m}_{g(i)j}} \mbox{X}_{i} (Y_{ij}- \widetilde{m}_{g(i)j} - \mbox{X}_{i}\mbox{\boldmath $ \alpha $}_{j} ) + \sum_{i: Y_{ij}< \widetilde{m}_{g(i)j}} \mbox{X}_{i} (\widetilde{m}_{g(i)j} - Y_{ij} - \mbox{X}_{i}\mbox{\boldmath $ \alpha $}_{j} ) \right\} \nonumber \\
&=& \frac{1}{\sqrt{n}} \left\{ \sum_{i: Y_{ij}\geq \widetilde{m}_{g(i)j}} \mbox{X}_{i} (Y_{ij}- m_{g(i)j} - \mbox{X}_{i}\mbox{\boldmath $ \alpha $}_{j} ) + \sum_{i: Y_{ij}< \widetilde{m}_{g(i)j}} \mbox{X}_{i} ( m_{g(i)j} - Y_{ij} - \mbox{X}_{i}\mbox{\boldmath $ \alpha $}_{j} ) \right\} \label{var1} \\
  & & + \frac{1}{\sqrt{n}} \sum_i \mbox{X}_{i} (\widetilde{m}_{g(i)j} - m_{g(i)j}) \left\{ I(Y_{ij} < \widetilde{m}_{g(i)j}) - I(Y_{ij} \geq \widetilde{m}_{g(i)j}) \right\}. \label{var2}
\end{eqnarray}
The first component (\ref{var1}) is the usual score function as if $m_{g(i)j}$ is known, and the second component is the extra variability due to estimation of the medians. The asymptotic distribution of the sample median has been well established to be a normal distribution \citep{VanderVaart2005}. It maximizes the objective function $- \sum_{g(i)} |Y_{ij} - m_{g(i)j}|$, with the following asymptotic linear expansion
\begin{eqnarray}
\sqrt{n_g} (\widetilde{m}_{g(i)j} - m_{g(i)j}) &=&  -\frac{1}{2f(m_{g(i)j})} \frac{1}{\sqrt{n_g}} \sum  -\mbox{sign}(Y_{ij} - m_{g(i)j})  + o_p(1) \nonumber \\
&\rightarrow_d& \mathcal{N}\left(0, \frac{1}{\{2f(m_{g(i)j})\}^2}\right), \nonumber
\end{eqnarray}
where $n_g$ is sample size for group $g$ and $f(m_{g(i)j})$ is the probability density of $Y_{ij}$ at its group median. Using these derivations, the sandwich covariance matrix of ($\widehat{\mbox{\boldmath $ \beta $}}_{j}$, $\widehat{\mbox{\boldmath $ \alpha $}}_{j}$) as expressed in (\ref{asymp}) can be modified to account for the estimation of the group sample medians. In {\bf R} package \texttt{DMtest}, we implemented kernel density estimation for $f(m_{g(i)j})$, which appears to be satisfactory in simulation studies (the corrected 2-df test in Table 1).

\subsection{Constrained hypothesis incorporating hypervariable cancer methylation}


As shown in Figure 1, the cancer CpG methylation is almost always more variable than the normal CpG methylation. The standard 2-df hypothesis testing using H$_{1a}$ may not be optimal in power. Two constrained alternative hypotheses can be constructed:
\begin{equation}
\mbox{H}_{0}: \beta_{1j}=0, \alpha_{1j}=0  \mbox{\hspace{10pt} versus \hspace{10pt}}\mbox{H}_{1b}: \beta_{1j} \neq 0 , \alpha_{1j} \geq 0. \label{hypothesis2}
\end{equation}
\begin{equation}
\mbox{H}_{0}: \beta_{1j}=0, \alpha_{1j}=0  \mbox{\hspace{10pt} versus \hspace{10pt}}\mbox{H}_{1c}: \beta_{1j} \geq 0 , \alpha_{1j} \geq 0. \label{hypothesis3}
\end{equation}
H$_{1b}$ tests for the differential mean methylation and the increased variability in cancer samples, reducing the parameter space under the alternative hypothesis by half. H$_{1c}$ further restricts the parameter space, testing for increased mean and increased variability in tumor samples. This is motivated by biomarker studies that specifically detect CpGs with low or no methylation in normal samples and increased methylation for cancer samples.

The one-sided likelihood ratio test (LRT) for the constrained hypothesis (\ref{hypothesis3}) has been studied over decades in the statistical literature, mostly for clinical trials with multiple study endpoints\citep{Kudo63,Perlman69}. The asymptotic null distribution of the one-sided statistics are generally difficult to obtain, and maximizing the likelihood under the one-sided constraints can be strenuous for genome-wide testing. Because the constraint is only on two parameters, we were able to develop an efficient algorithm to compute the LRT under the constrained hypotheses H$_{1b}$ and H$_{1c}$, as implemented in the \texttt{DMVC} function in the {\bf R} package \texttt{DMtest}.

To test the constrained hypothesis $\mbox{H}_{1b}$ or $\mbox{H}_{1c}$, the likelihood ratio test (LRT) can be conducted by  treating ($\hat{\beta}_{1j}$, $\hat{\alpha}_{1j}$) as a pair of data points from its estimated bivariate asymptotic normal distribution. Specifically, the LRT statistic is the difference of the log likelihood under the null hypothesis ($\alpha_{1j}=\beta_{1j}=0$) and the log likelihood under the constrained alternative hypothesis (either $\mbox{H}_{1b}$ or $\mbox{H}_{1c}$), expressed as
\[
(\hat{\beta}_{1j}, \hat{\alpha}_{1j})^T\widehat{\Sigma}^{-1}(\hat{\beta}_{1j}, \hat{\alpha}_{1j}) - (\hat{\beta}_{1j} -\tilde{\beta}_{1j}, \hat{\alpha}_{1j}-\tilde{\alpha}_{1j})^T\widehat{\Sigma}^{-1}(\hat{\beta}_{1j}-\tilde{\beta}_{1j}, \hat{\alpha}_{1j}-\tilde{\alpha}_{1j}),
\]
where $\tilde{\beta}_{1j}$ and $\tilde{\alpha}_{1j}$ are maximal likelihood estimates under the constrained parameter space. Because of the two dimensional parameter space and the simplicity of this likelihood, an algebraic solution for the constrained MLE under H$_{1b}$ or H$_{1c}$ can be obtained as follows. If $(\hat{\beta}_{1j}, \hat{\alpha}_{1j})$ falls in the constrained parameter space, then the likelihood ratio statistic reduces to $(\hat{\beta}_{1j}, \hat{\alpha}_{1j})^T\widehat{\Sigma}^{-1}(\hat{\beta}_{1j}, \hat{\alpha}_{1j})$. If $(\hat{\beta}_{1j}, \hat{\alpha}_{1j})$ falls out of the constrained parameter space,  since $(\hat{\beta}_{1j} -\tilde{\beta}_{1j}, \hat{\alpha}_{1j}-\tilde{\alpha}_{1j})^T\widehat{\Sigma}^{-1}(\hat{\beta}_{1j}-\tilde{\beta}_{1j}, \hat{\alpha}_{1j}-\tilde{\alpha}_{1j})$ is a convex function, the minimizer in the constrained parameter space is achieved on the boundary.
\begin{itemize}
\item For H$_{1b}$, the boundary is $\alpha_{1j}= 0$; The corresponding minimizer for H$_{1b}$ can be found  on  $\beta_{1j}= 0$, $\alpha_{1j}\geq 0$, by solving a quadratic function.

\item For H$_{1c}$, the boundary is $\beta_{1j}\geq 0$, $\alpha_j =0$ and $\beta_{1j}= 0$, $\alpha_j \geq 0$, namely the non-negative axis of $\beta_{1j}$ and $\alpha_{1j}$. For H$_{1c}$, it is sufficient to first find the local minimizer in  $\beta_{1j}\geq 0$, $\alpha_{1j}= 0$ and the local minimizer in $\beta_{1j}= 0$, $\alpha_{1j}\geq 0$. The constrained global MLE is the minima with the smaller value of the objective function.
\end{itemize}
Compared to the standard numerical algorithm for constrained maximization, this algebraic solution drastically reduces computation time and enables genome-wide CpG testing.

\begin{figure}[!ht]
   \includegraphics[width=.45\textwidth]{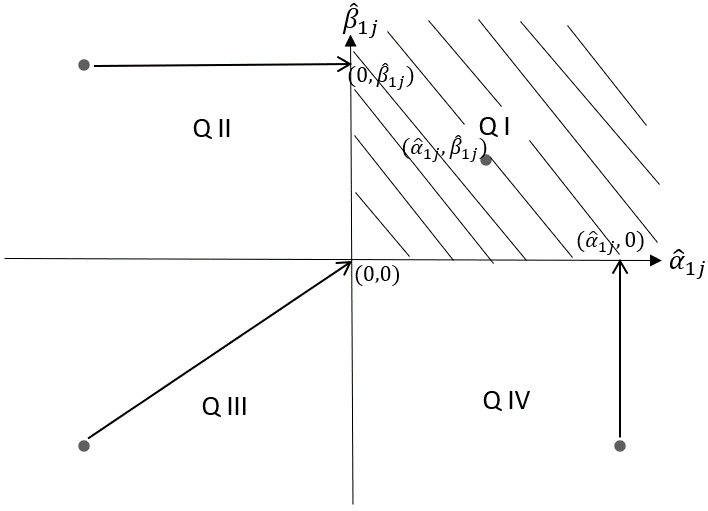}
   \centering
   \caption{Constrained Hypothesis Test with Identity Covariance}
   \label{one_sided_figure}
\end{figure}

The LRT statistics for testing constrained hypotheses typically take a Chi-Bar-square distribution. Chapter 3 of \cite{Silvapulle2005} gives the general theory for the constrained hypothesis testing for multivariate normal mean. To illustrate, let us assume for simplicity the asymptotic covariance ${\Sigma}$ is the identity matrix. As shown in Figure 1 for the constrained alternative hypothesis H$_{1c}$, the signs of $\hat{\beta}_{1j}$ and $\hat{\alpha}_{1j}$ determine the constrained MLEs, ($\tilde{\beta}_{1j}, \tilde{\alpha}_{1j}$). Specifically, if $\hat{\beta}_{1j}$ and $\hat{\alpha}_{1j}$ fall in the first quadrant, then  $\tilde{\beta}_{1j} =\hat{\beta}_{1j}$, $ \tilde{\alpha}_{1j} =\hat{\alpha}_{1j} $; if $\hat{\beta}_{1j}$ and $\hat{\alpha}_{1j}$ fall in the second quadrant, $\tilde{\beta}_{1j} = \hat{\beta}_{1j}$, $\tilde{\alpha}_{1j} = 0$; if $\hat{\beta}_{1j}$ and $\hat{\alpha}_{1j}$ fall in the third quadrant, $\tilde{\beta}_{1j} = 0$, $\tilde{\alpha}_{1j} = 0$; if $\hat{\beta}_{1j}$ and $\hat{\alpha}_{1j}$ fall in the fourth quadrant, $\tilde{\beta}_{1j} = 0$, $\tilde{\alpha}_{1j} =\hat{\alpha}_{1j}$. Therefore the LRT statistic becomes,
\[
\text{LRT}=\begin{cases}
         \hat{\alpha}_{1j}^2+\hat{\beta}_{1j}^2  & \text{if ($ \hat{\alpha}_{1j}, \hat{\beta}_{1j}$) in Q I;} \\
        \hat{\beta}_{1j}^2  & \text{if ($ \hat{\alpha}_{1j}, \hat{\beta}_{1j}$) in Q II;} \\
       0 & \text{if ($ \hat{\alpha}_{1j}, \hat{\beta}_{1j}$) in Q III;} \\
\hat{\alpha}_{1j}^2 & \text{if ($ \hat{\alpha}_{1j}, \hat{\beta}_{1j}$) in Q IV.}
              \end{cases}
\]
\noindent Corresponding to the four quadrants, the null distribution of LRT is therefore a mixture of $\chi_2^2, \chi_1^2, \chi_0^2 \mbox{ (a point mass)}$.  When the asymptotic variance matrix $\Sigma_j$ is known (typically not an identity matrix), a geometric rotation will conclude that the null distribution for this LRT statistic under constrained hypothesis testing takes a Chi-Bar-square distribution\citep{Silvapulle2005}, namely
\[
\mbox{Pr}(\mbox{LRT}_j\leq c| \mbox{H}_0) = q \mbox{Pr}(\chi_0^2 \leq c) + 0.5 \mbox{Pr}(\chi_1^2 \leq c) + (0.5-q) \mbox{Pr}(\chi_2^2 \leq c),
\]
where $q = (2\pi)^{-1} \cos^{-1}(\rho_j)$, $\rho_j$ is the correlation between $\hat{\beta}_{1j}$ and $\hat{\alpha}_{1j}$ for the $j^{th}$ marker. For the constrained hypothesis H$_{1b}$, the distribution of LRT is much simpler because the constraint is on one parameter only. Depending on where $\hat{\beta}_{1j}$ and $\hat{\alpha}_{1j}$ falls,  the LRT is either a $\chi_2^2$ or $\chi_1^2$ distribution  \citep{Silvapulle2005}. It can be shown that the null distribution for H$_{1b}$ can be expressed as
\[
\mbox{Pr}(\mbox{LRT}_j\leq c| H_0) =0.5 \mbox{Pr}(\chi_1^2 \leq c) + 0.5 \mbox{Pr}(\chi_2^2 \leq c).
\]

\subsection{Comparison to pAUC for biomarker discovery}

To capture tumor heterogeneity when comparing gene expressions between cancer and normal samples, it was proposed to use pAUC as the discriminatory measure to select genes from microarray experiments \citep{Pepe2003}.  In our notation, let $Y_{ij}^C$ denote methylation values for normal (control) samples and  $Y_{ij}^D$ denote methylation values for disease (cancer) samples.
Let $F_C$ and $F_D$ denote the corresponding cumulative distribution functions, and $n_C$ and $n_D$ denote the corresponding sample sizes. As shown in Figure 3, the ROC curve characterizes the separation of distributions of $Y_{ij}^C$  and $Y_{ij}^D$, and the pAUC focuses on the upper quantile range of normal sample values. Note that pAUC can be also considered one-sided metric, because it targets at good sensitivity for high specificity values, assuming the marker values for cancer cases are more likely to be greater than those in normal cases.
\begin{eqnarray}
\mbox{ROC}(t) &=& 1- F_D \{F_C^{-1}(1-t)\}, \nonumber \\
\mbox{pAUC}(t) &=& \int_0^t \mbox{ROC}(t) dt \nonumber .
\end{eqnarray}
Cancer early-detection biomarkers often require a high specificity (low false positive rate) and a clinically meaningful sensitivity, the area in the ROC curve captured by pAUC.

\begin{figure}[!ht]
   \includegraphics[width=.45\textwidth]{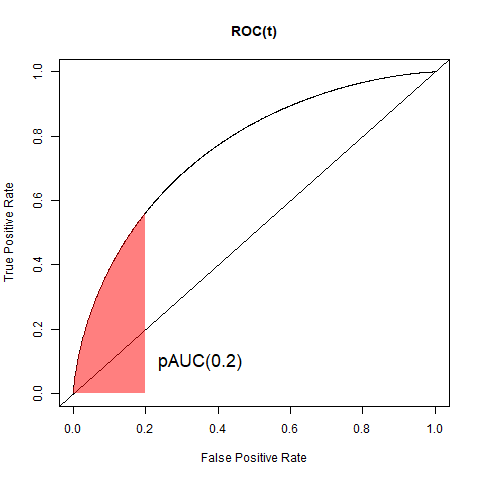}
   \centering
   \caption{ROC and pAUC}
   \label{pAUC}
\end{figure}

While it is informative to use pAUC to rank genes, as pAUC is arguably the most clinically relevant measure, it is critical to develop a corresponding hypothesis testing procedure that can gauge the significance level of an observed pAUC when testing tens of thousands of genes simultaneously. Under the null hypothesis that there is no difference between cancer and normal samples,  pAUC($t_0$)=$\frac{t_0^2}{2}$, where $t_0$ is the target false positive rate. The nonparametric estimate of pAUC($t_0$) is expressed as
\begin{equation}
\mbox{p}\widehat{\mbox{AUC}}(t_0) = \frac{\sum_i \sum_j \delta_i \delta_j I(Y_{ij}^D > Y_{ij}^C, Y_{ij}^C > \hat{F}^{-1}_C(1-t_0))}{\sum_i \sum_j \delta_i \delta_j}, \nonumber
\end{equation}
where $\delta_i$ and $\delta_j$ are point masses of observed values in cancer and normal samples.  Note that the nonparametric estimate of pAUC($t_0$) is a sum of indicator functions that is evaluated in the high-specificity area of the data, therefore can be variable for small sample sizes.  Assume $n_D/n \rightarrow \lambda \in (0,1)$ as the sample size $n \rightarrow \infty$.  By Theorem 3 in Wang and Huang \citep{Wang2019},  as $n \rightarrow \infty, \sqrt{n}[\mbox{p}\widehat{\mbox{AUC}}(t_0)-\mbox{pAUC}(t_0)]$ converges to a normal random variable with mean zero and variance,
\begin{align*}
    \sigma_{\mbox{pAUC}(t_0)}^2 = & \frac{1}{\lambda}\mbox{Var}(J_D)+ \frac{1}{1-\lambda}\mbox{Var}(J_C) \\
    &+ \frac{1}{1-\lambda}\Big\{[1-F_D(q_0)]^2\mbox{Var}\big[H_C(q_0)\big]+2[1-F_D(q_0)]\mbox{Cov}\big[J_D,H_C(q_0)\big]\Big\},
\end{align*}
where $q_0=F_C^{-1}(1-t_0), J_D=\mbox{Pr}\big(Y^C<Y^D,Y^C\in (q_0,\infty)|Y^C\big), J_C=\mbox{Pr}\big(Y^D>Y^C,Y^C\in (q_0,\infty)|Y^D\big)$ and $H_C(q)=I(Y^C<q)$.

We used the large-sample variance estimate provided above to compute a p-value. One challenge for the nonparametric pAUC estimate in high-dimensional hypothesis testing is that its p-value needs to be accurate even at the extremely small level, e.g., $10^{-7}$ or lower. This requires that the point estimate and the asymptotic variance work well even in small samples. However, if the target $t_0$ is 0.05 (a relatively high type I error rate), pAUC evaluation restricts to the upper 5\% quantile range of the normal sample values, which amounts 2$\sim$3 samples if the total number of samples is 50, potentially leading to unstable estimates. In typical biomarker discovery studies, the number of normal or control samples may be small. The tail of the distribution for the nonparametric estimates of pAUC may not be approximated by the asymptotic distribution. We will show next the poor finite sample behavior of pAUC in simulation experiments (Table 1 and Figure 4).


\section{Results}
\label{sec3}

\subsection{Simulation study}

Seven methods were compared in simulated datasets: the two-sample $t$-test with unequal variances, Levene test for equal variances, the test for pAUC(0.2), the standard 2-df test for equal means and equal variances (2-df naive), the 2-df test accounting for variability of sample medians (2-df corrected), and the two proposed constrained joint tests for H$_{1b}$ and H$_{1c}$. To evaluate the performance of controlling the false positive rate, four probabilistic models under the null hypothesis were generated: a normal distribution $\mathcal{N}(0,1)$, a beta(10,90) distribution with mean 0.1, a chi-square distribution with three degrees of freedom $\chi^2_3$, a beta(10,90) distribution with 5\% outliers equally distributed across cancer and control samples (Beta$^+$ in Table \ref{type_I_error_simulations}). Equal numbers of cancer and control samples were generated, and the number of total samples increase from 50 (half cases and half controls) to 100, 250, and 500. The type I error rates at $p=$0.05 and $p=$0.001 were evaluated in 50,000 simulated datasets.

\begin{figure}[!ht]
\centering
 \includegraphics[width=.45\textwidth,height=7cm]{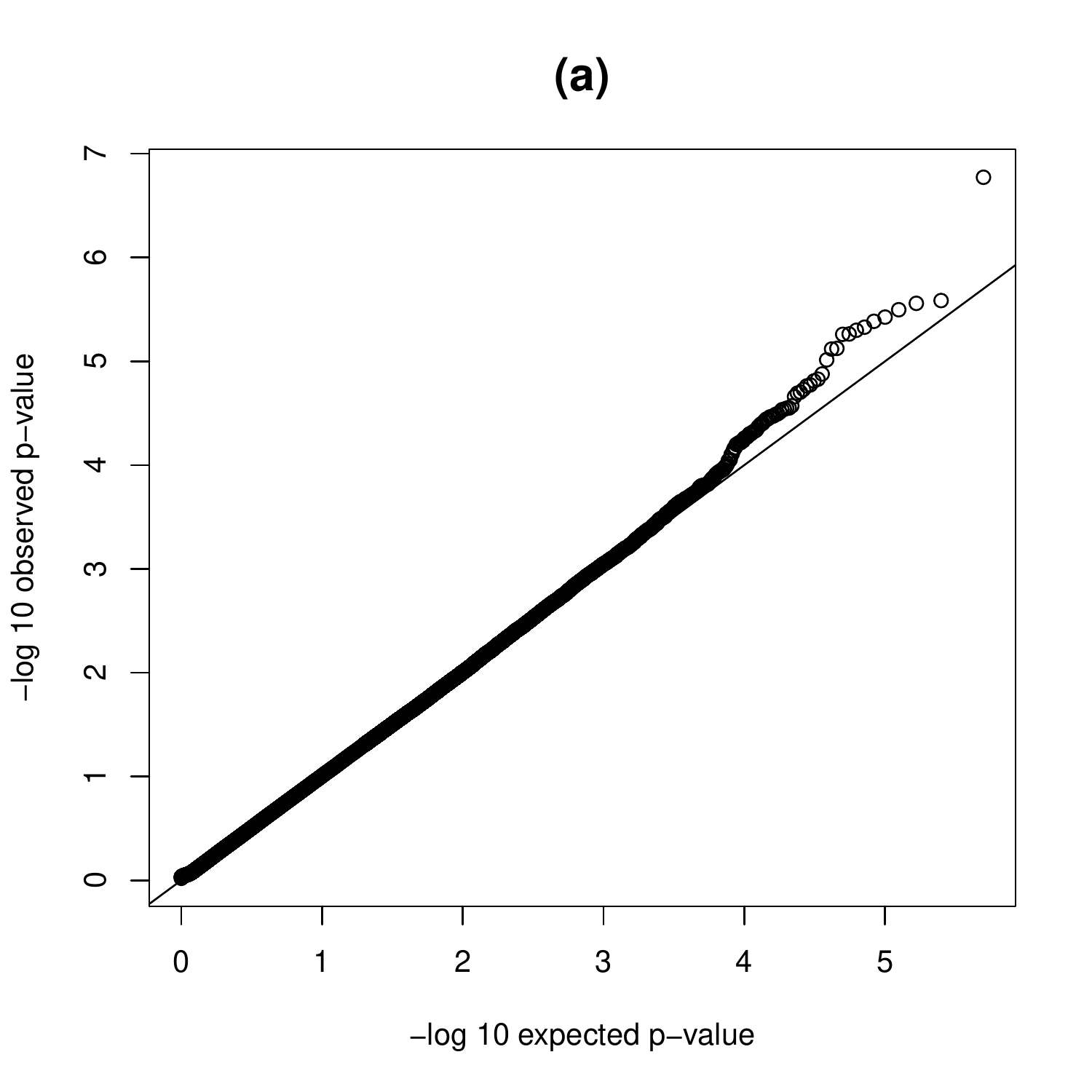}
   \includegraphics[width=.45\textwidth,height=7cm]{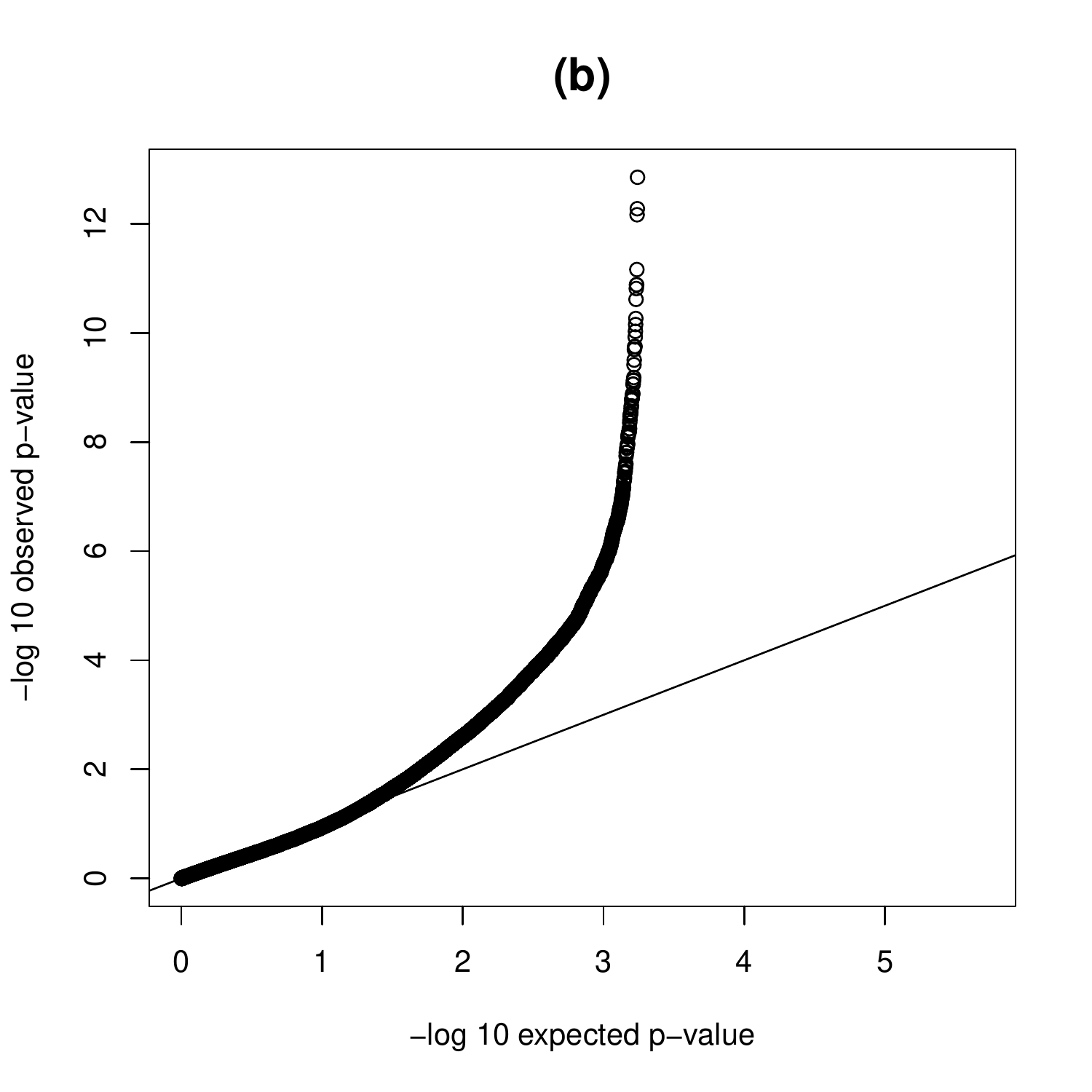}\\
     \includegraphics[width=.45\textwidth,height=7cm]{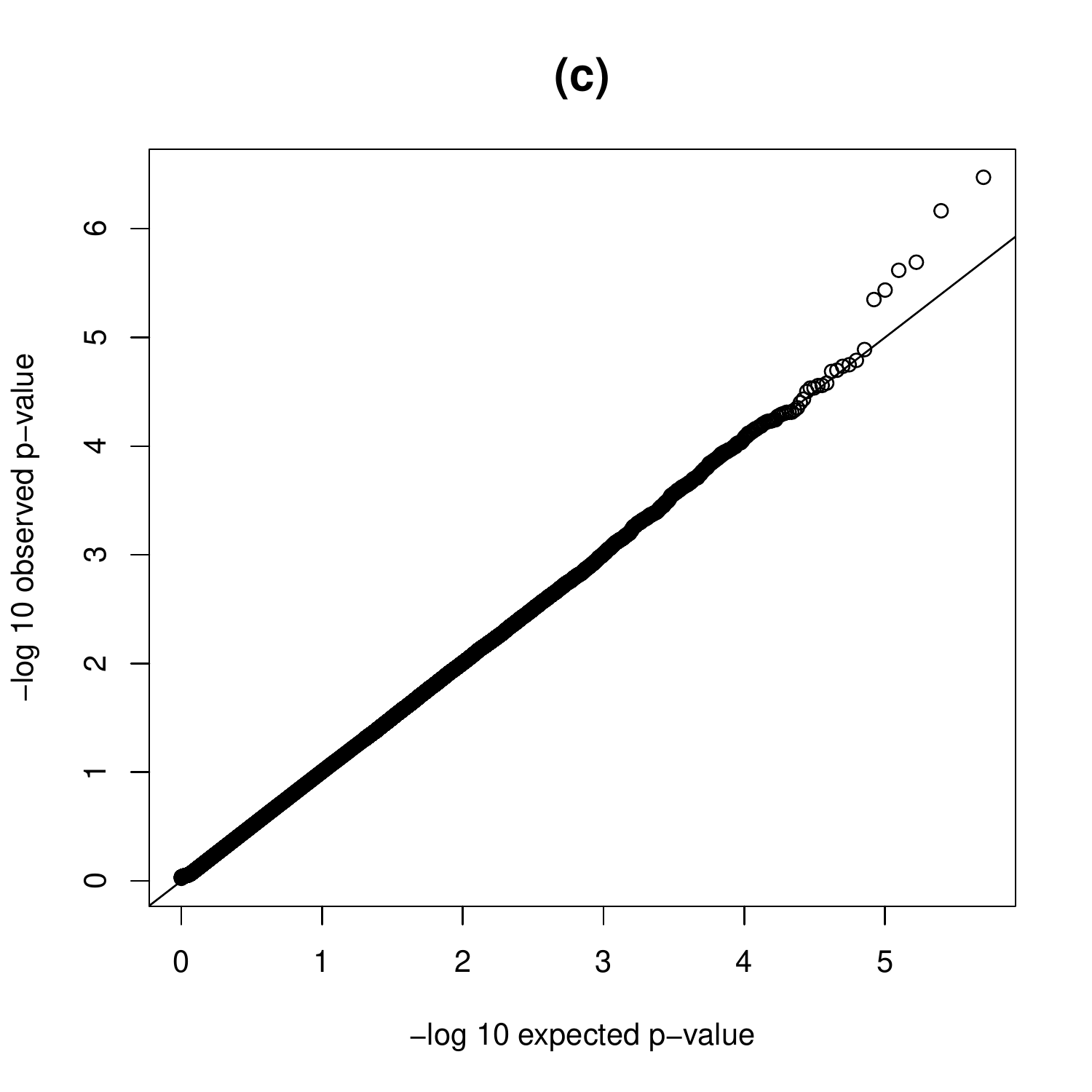}
       \includegraphics[width=.45\textwidth,height=7cm]{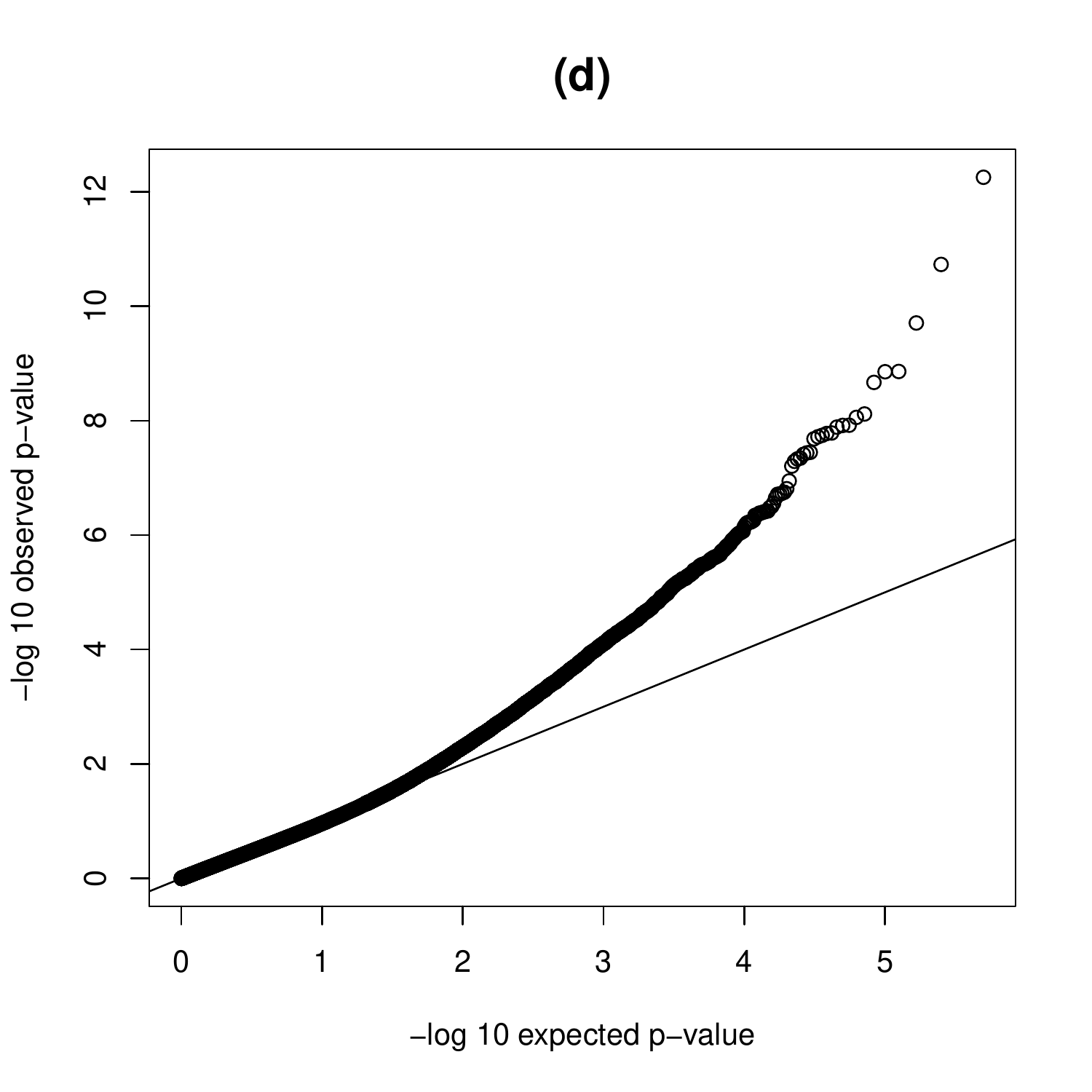}\\
     \includegraphics[width=.45\textwidth,height=7cm]{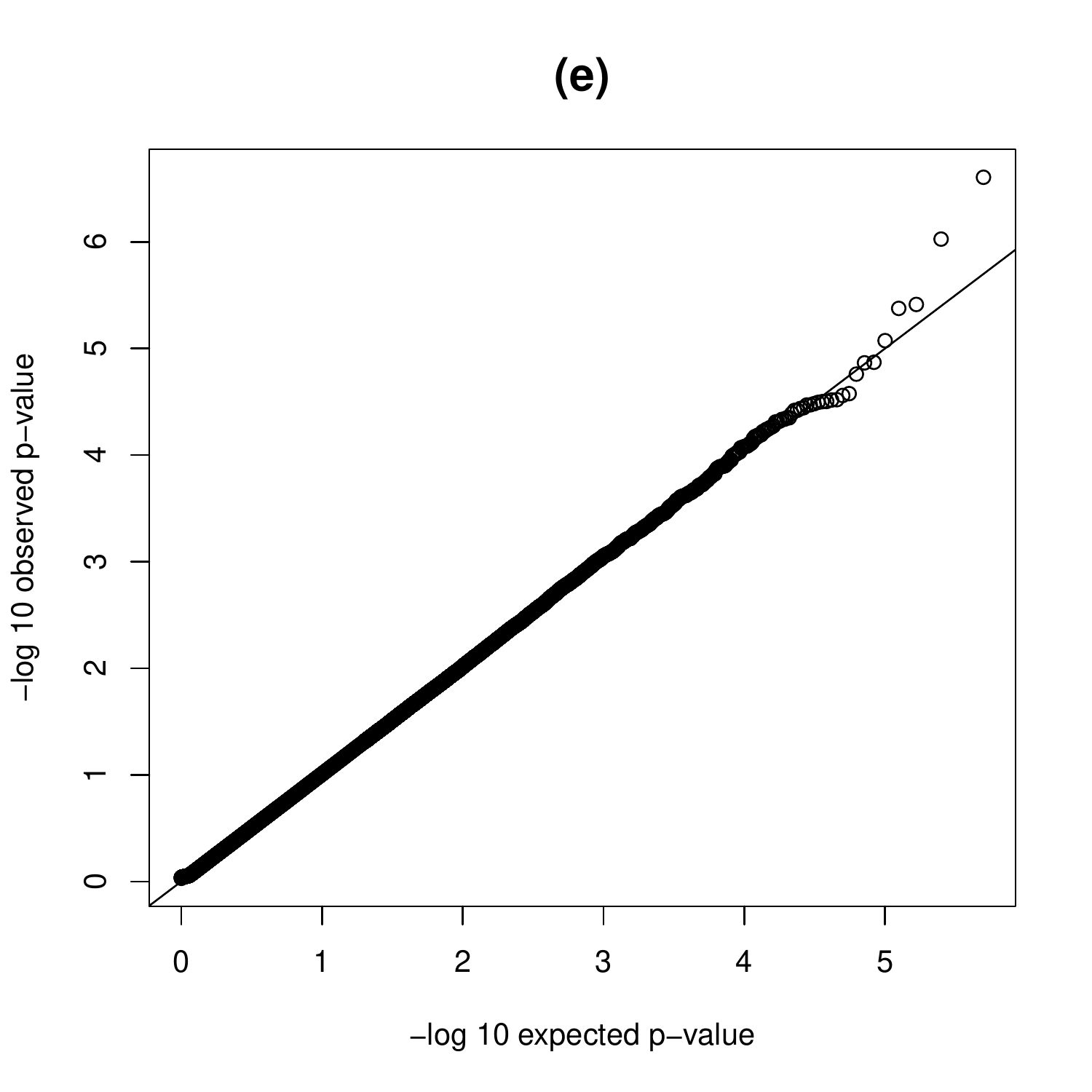}
       \includegraphics[width=.45\textwidth,height=7cm]{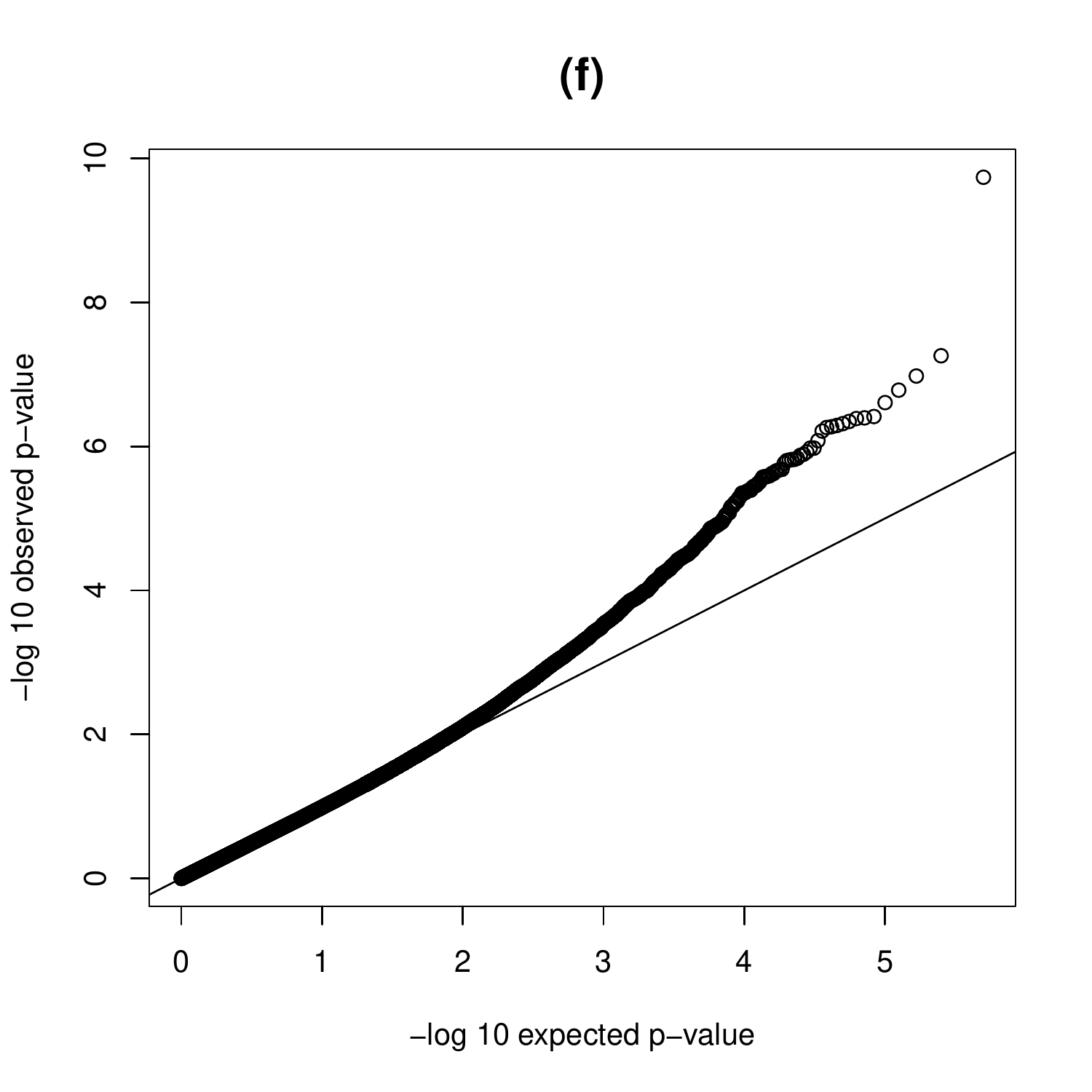}
        \caption{A comparison of q-q plots for $5 \times 10^5$ markers under different sample sizes. (a) 50 cases and 50 controls, DMVC$^{+}$; (b) 50 cases and 50 controls, pAUC;  (c) 100 cases and 100 controls, DMVC$^{+}$; (d) 100 cases and 100 controls, pAUC; (c) 250 cases and 250 controls, DMVC$^{+}$; (d) 250 cases and 250 controls, pAUC.   }
\label{qqplot_simulations}
      \end{figure}

Table \ref{type_I_error_simulations} shows the empirical type I error rates for 4 null models with increasing sample sizes from 25 per group to 250 per group. For the p-value cut-off 0.05, all seven methods except the naive 2-df test perform well and appear to properly control the false positive rate when sample size gets to at least 50 cases and 50 controls. The naive 2-df test has an inflated type I error rate when n=25 per group, for normal, beta and chi-square distributions (in red color). Its performance improves when n increases. However, when the p-value cut-off decreases to 0.001, the observed false positive rates for pAUC and the naive 2-df test appear to deteriorate rapidly, reaching as high as 0.0097 for pAUC (nearly 10 fold inflation) and 0.005 for the naive 2-df test (5 fold inflation). Their performance are improved with the increasing sample size, though pAUC still have a minor degree of inflation even when there are 250 cases and 250 controls. This could be a serious problem for use of pAUC in high-dimensional testing, as we further show next in Figure \ref{qqplot_simulations}. The 2-df test correcting for estimation of sample medians has a substantially improved performance, particularly when sample size is small. All other methods include the two proposed constrained tests yield a much better performance in controlling the type I error rate at p=0.001.

\begin{sidewaystable}

\caption{\label{type_I_error_simulations} Empirical type I error rates of the proposed tests and comparator tests in 50,000 simulation datasets, when normal p-value cut-off is set to be 0.05 or 0.001.}
\centering
      {\begin{tabular}{@{}lllccccccc@{}}
        \hline
        &  &    &  &  & 2-df (H$_{1a}$) & 2-df (H$_{1a}$)  & pAUC &    Constrained   &       Constrained \\
& & n & t  & Levene   & naive & corrected &(0.2) & H$_{1b}$ &  H$_{1c}$   \\ \hline
p-value=0.05& Normal  & 25/25 &   0.0493 &0.0481 & \textcolor{red}{0.0751} &0.0463 &0.0593 &0.0521 &0.0474\\
& & 50/50 &    0.0492 &0.0495 &\textcolor{red}{0.0601} &0.0462 &0.0513 &0.0494 &0.0484\\
& &  100/100 &  0.0506 &0.0473 &0.0540 &0.0470 &0.0482 &0.0497 &0.0486\\
& & 250/250 & 0.0495 &0.0482 &0.0509 &0.0481 &0.0480 &0.0493 &0.0478\\
&Beta  & 25/25 &   0.0499 &0.0494 &\textcolor{red}{0.0740} &0.0463 &0.0592 &0.0520 &0.0469\\
&   &50/50 &   0.0496 &0.0497 &\textcolor{red}{0.0600} &0.0467 &0.0522 &0.0499 &0.0490\\
&   &100/100 & 0.0513 &0.0487 &0.0541 &0.0470 &0.0475 &0.0499 &0.0475\\
&   &250/250 & 0.0499 &0.0491 &0.0512 &0.0487 &0.0474 &0.0492 &0.0475\\
&Chisq &  25/25 &  0.0452 &0.0507 &\textcolor{red}{0.0672} &0.0380 &0.0583 &0.0405 &0.0402\\
&   &50/50 & 0.0483 &0.0519 &0.0580 &0.0427 &0.0512 &0.0445 &0.0447\\
&   &100/100 & 0.0507 &0.0512 &0.0536 &0.0463 &0.0491 &0.0478 &0.0471\\
&   &250/250 & 0.0498 &0.0506 &0.0512 &0.0483 &0.0484 &0.0493 &0.0495\\
&Beta$^{+}$ & 25/25 &0.0294 &0.0290 &0.0557 &0.0358 &0.0568 &0.0415 &0.0413\\
&   &50/50 &  0.0437 &0.0464 &0.054 &0.0434 &0.0448 &0.0465 &0.0469\\
&   &100/100 & 0.0480 &0.0493 &0.0521 &0.0471 &0.0420 &0.0480 &0.0484\\
&   &250/250 &  0.0507 &0.0504 &0.0512 &0.0495 &0.0447 &0.0487 &0.0490\\ \hline
p-value=0.001 &  Normal  & 25/25 & 0.0010 &7e-04 &\textcolor{red}{0.0050} &0.0021 &\textcolor{red}{0.0089} &0.0024 &0.0018\\
& & 50/50 &  7e-04 &9e-04 &\textcolor{red}{0.0021} &0.0012 &\textcolor{red}{0.0056} &0.0014 &0.0011\\
& &  100/100 & 9e-04 &9e-04 &0.0016 &0.0012 &\textcolor{red}{0.0034} &0.0011 &0.0011\\
& &  250/250 &  0.0011 &9e-04 &0.0011 &9e-04 &\textcolor{red}{0.0020} &0.0010 &9e-04\\
&Beta  & 25/25 &  0.0010 &7e-04 &\textcolor{red}{0.0046} &0.0019 &\textcolor{red}{0.0097} &0.0022 &0.0018\\
&   &50/50 &  8e-04 &0.0010 &\textcolor{red}{0.0022} &0.0011 &\textcolor{red}{0.0061} &0.0014 &0.0012\\
&   &100/100 &0.0011 &6e-04 &0.0015 &0.0011 &\textcolor{red}{0.0033} &0.0011 &9e-04\\
&   &250/250 & 9e-04 &0.0012 &0.0013 &0.0011 &\textcolor{red}{0.0021} &0.001 &0.0010\\
&Chisq & 25/25 &  7e-04 &0.0011 &\textcolor{red}{0.0035} &0.0012 &\textcolor{red}{0.0088} &0.0013 &0.0013\\
&   &50/50 & 6e-04 &0.0011 &\textcolor{red}{0.0023} &0.0013 &\textcolor{red}{0.0055} &0.0013 &0.0012\\
&   &100/100 & 0.0011 &0.0011 &0.0016 &0.0011 &\textcolor{red}{0.0039} &0.0011 &0.0012\\
&   &250/250 & 0.0010 &0.0011 &0.0011 &0.0010 &\textcolor{red}{0.0027} &0.0012 &0.0011\\
&Beta$^{+}$ &  25/25 & 1e-04 &1e-04 &\textcolor{red}{0.0025} &0.0011 &\textcolor{red}{0.0087} &0.0012 &0.0011\\
&   &50/50 & 0.0001 &1e-04 &0.0010 &5e-04 &\textcolor{red}{0.0046} &7e-04 &7e-04\\
&   &100/100 & 0.0004 &4e-04 &0.0010 &8e-04 &\textcolor{red}{0.0029} &9e-04 &9e-04\\
&   &250/250 &0.0006 &6e-04 &9e-04 &8e-04 &0.0013 &0.0010 &9e-04\\
     \hline
      \end{tabular}}
\end{sidewaystable}

We further investigated the performance of pAUC and the proposed constrained test for H$_{1c}$ in genome-wide CpG testing with $5 \times 10^5$ markers. Figure \ref{qqplot_simulations} shows the q-q plots for pAUC and the proposed test when the distribution for all markers is a chi-square distribution with three degrees of freedom $\chi^2_3$. When sample size is 50 cases and 50 controls, the pAUC test shows a drastically inflated type I error rate for p-values smaller than 0.01, severely deviating from the diagonal line. The error control is improved with larger sample sizes, though even 250 cases and 250 controls do not yield a diagonal q-q line. These results prove the limitation of the nonparametric estimate of pAUC for high-dimensional testing, as we discuss in Section 2.3, particularly when sample sizes for a biomarker discovery study is small. As a comparison, the proposed constrained test delivered consistently well-behaved q-q lines in all three sample sizes (Figure \ref{qqplot_simulations}).

To evaluate the power of the proposed constrained method in a single hypothesis test, four scenarios were generated with 50 cases and 50 controls, each with some distributional differences between cancer samples and control samples. We included pAUC in this power comparison despite its problem in high-dimensional testing, because the p-value cut-off 0.05 for pAUC still yields a valid type I error rate. Methylation M values were generated by a Gaussian distribution or a mixture of Gaussian distributions. A parameter $\tau \in$ [0,0.7] was used to control the magnitude of the differences:  the first scenario is all cancer samples having both mean and variance increase, specifically $\mathcal{N}(\tau,1+\tau)$ vs $\mathcal{N}(0,1)$ ; the second scenario is all cancer samples having increased variance but no difference in mean, specifically $\mathcal{N}(0,1+\tau)$ vs $\mathcal{N}(0,1)$; the third scenario is all cancer samples having mean increase but no difference in variance , specifically $\mathcal{N}(\tau,1)$ vs $\mathcal{N}(0,1)$; the fourth scenario is 25\% cancer samples with increasing mean and variance, specifically a mixture of 75\% $\mathcal{N}(0,1)$ and 25\%  $\mathcal{N}(3\tau,1+3\tau)$.

\begin{figure}[!ht]

  \vspace{15pt}
\centering
   \includegraphics[width=.45\textwidth]{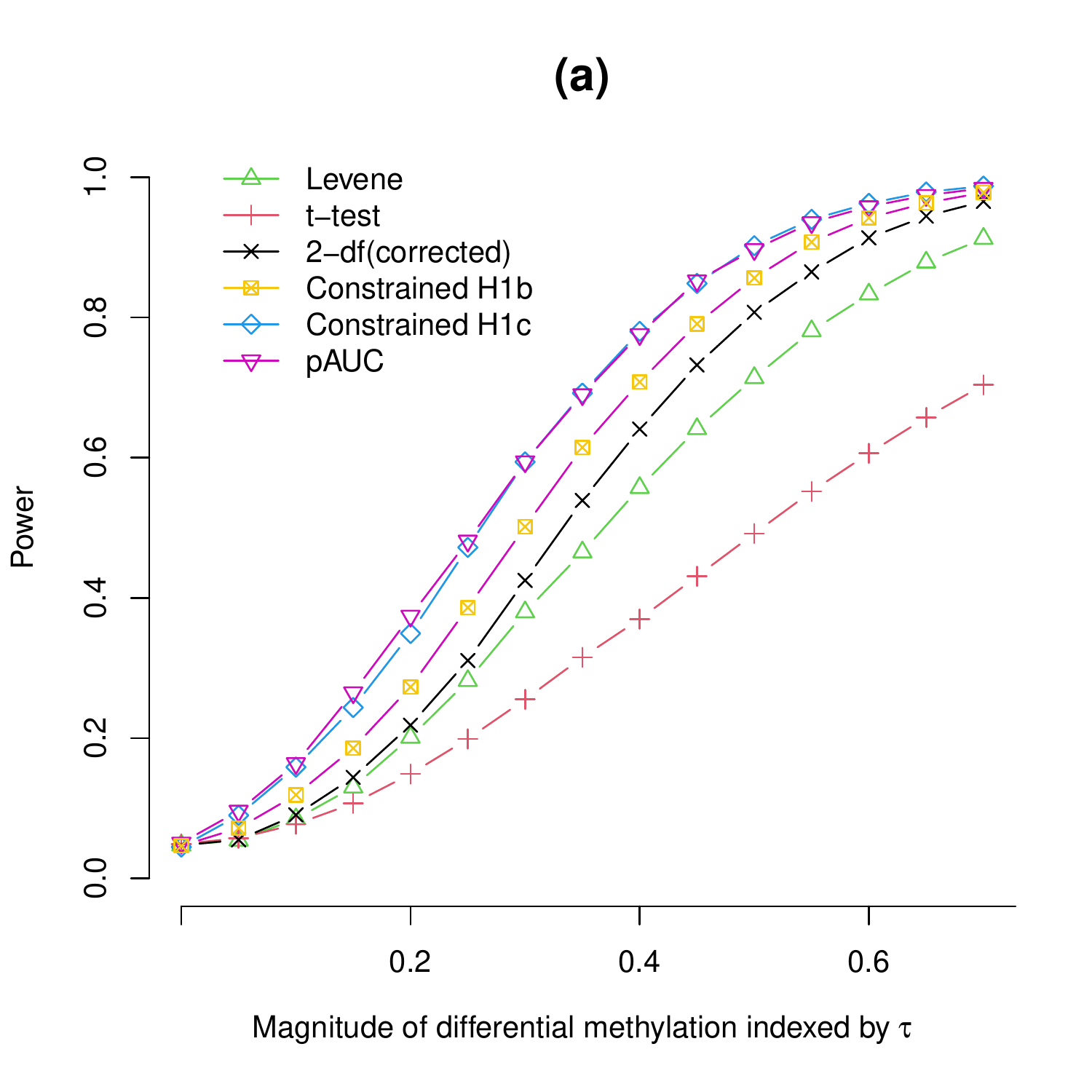}
      \includegraphics[width=.45\textwidth]{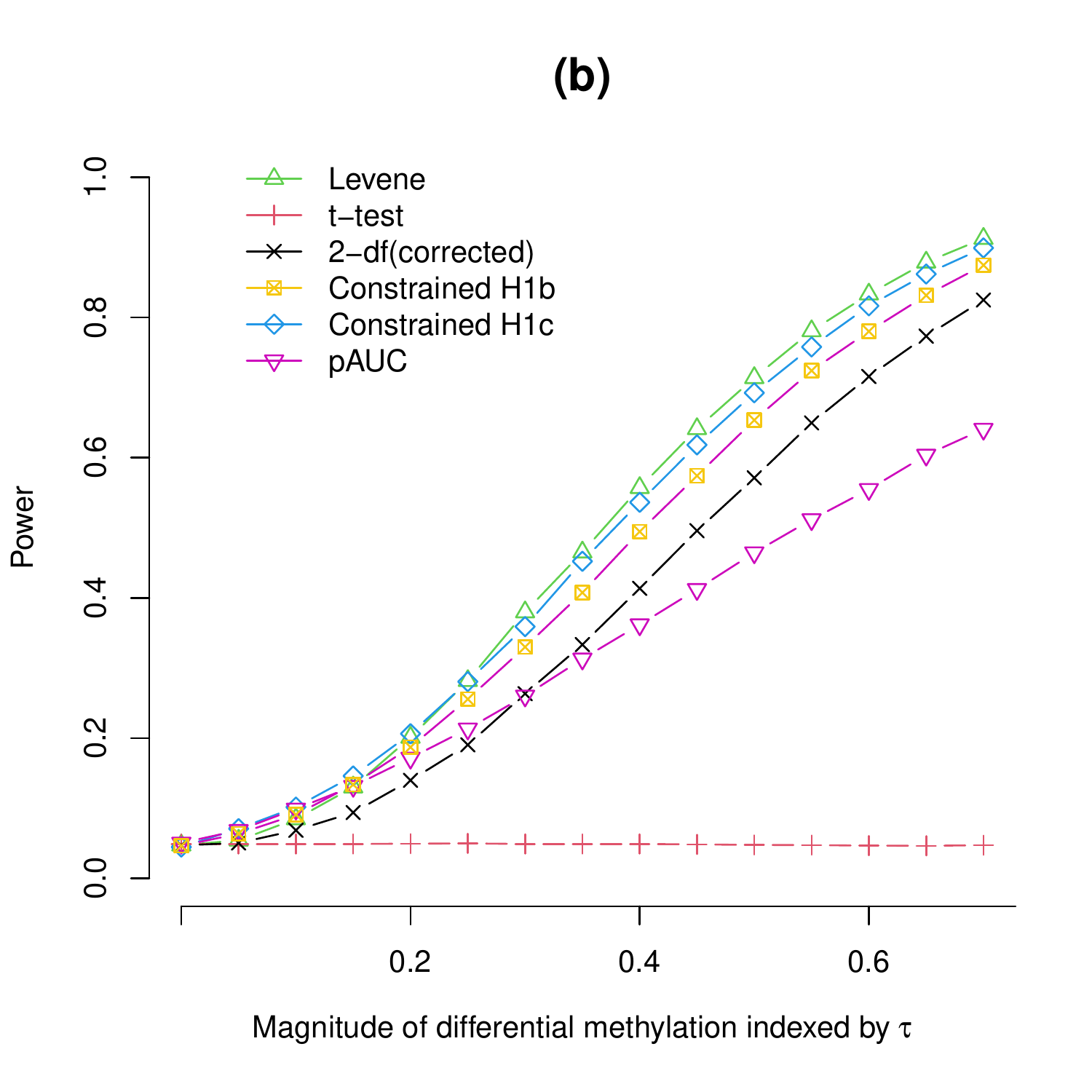}\\
       \includegraphics[width=.45\textwidth]{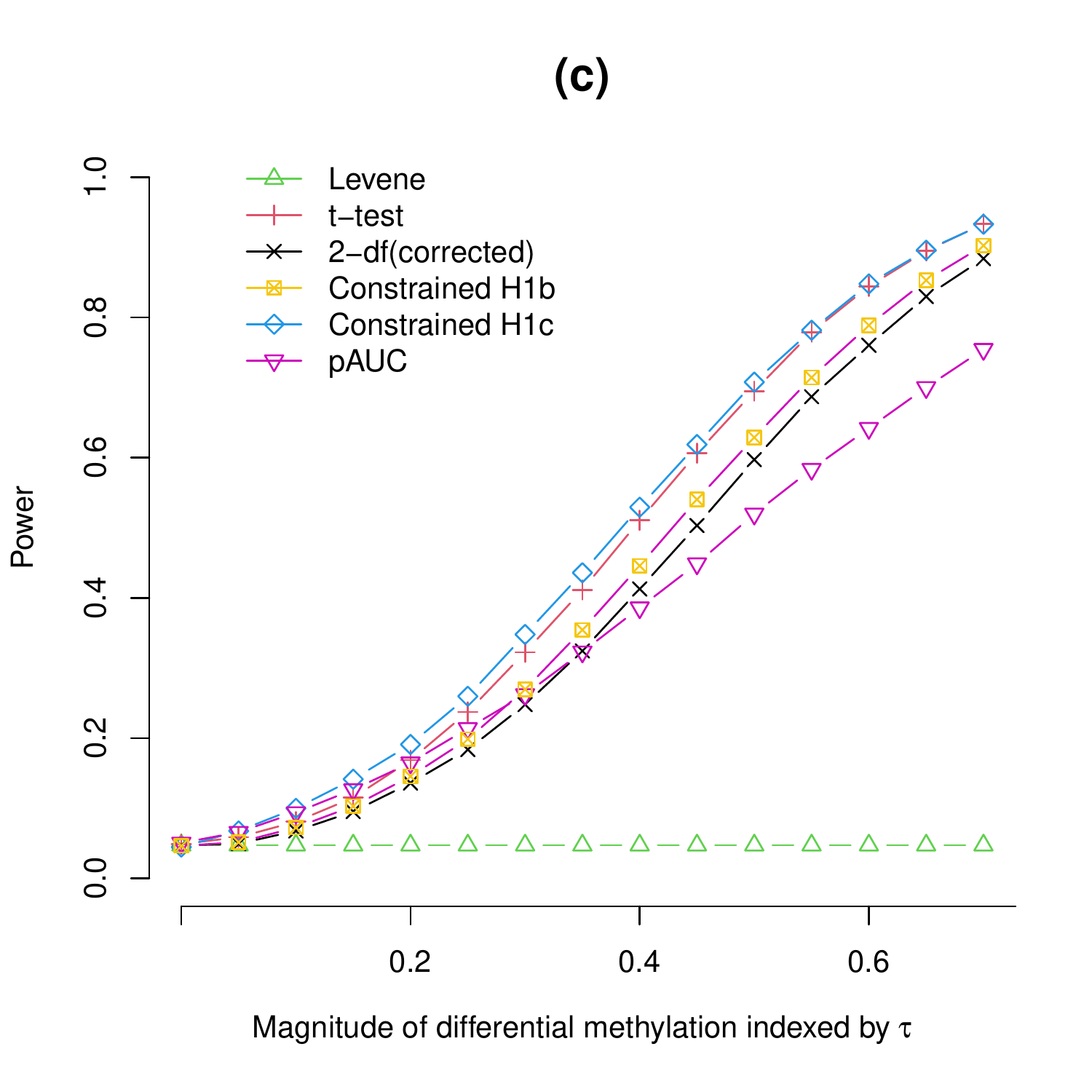}
      \includegraphics[width=.45\textwidth]{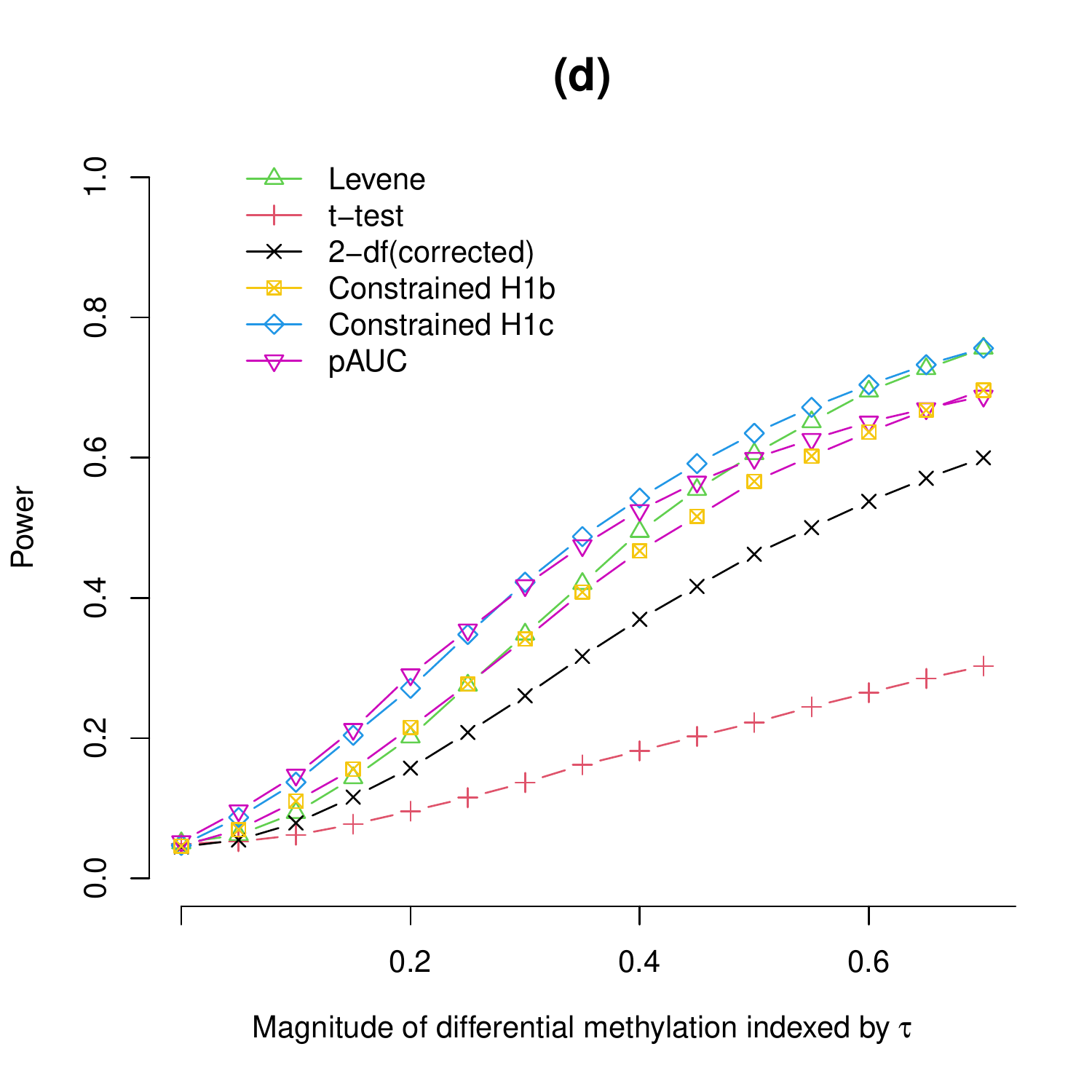}
        \caption{A comparison of power between the proposed joint constrained tests and the existing tests in simulated datasets. Benchmark methods include the Levene test, the t-test, and the 2-df test. (a) Cancer samples have increased mean and increased variance. (b)  Cancer samples have increased variance only. (c) Cancer samples have increased mean only. (d) A portion (25\%) of cancer samples have increased mean and increased variance. }
\label{power_plot_simulations}
      \end{figure}

Figure \ref{power_plot_simulations} shows the power to reject the null hypothesis (no differential methylation in mean nor variance) in 2,000 simulated datasets for the proposed constrained joint tests and the benchmark methods. In all scenarios the joint constrained test for $H_{1c}$ (increased mean and increased variance) delivered consistently the highest power, owing to its flexibility in detecting diverse signals (as compared to t-test or Levine test) and its focus on one-sided alternative hypotheses (as compared to the 2-df test). More restriction in the alternative parameter space increases the power, as seen from the comparison of the constrained tests for H$_{1b}$ and H$_{1c}$. The test for pAUC(0.2) performs competitively in the first scenario where there is both mean and variance increase, and in the fourth scenario where there is a subgroup with increased mean and variance. However when there is only increased mean or only increased variance, pAUC is clearly inferior to the proposed constrained test.

\subsection{Application to TCGA data}


Methylome data by the Infinium Methylation 450K assay for 2781 fresh-frozen tumor tissue samples were obtained from TCGA for 6 major cancers, namely prostate cancer (PCa),  colorectal cancer (CRC), lung squamous cell carcinoma (LUSC), and lung adenocarcinoma (LUAD),  breast cancer (BRCA), and liver hepatocellular carcinoma (HCC), and 308 matched adjacent normal samples. The sample sizes for each cancer and its matched normal samples are shown in Table \ref{all_DMresults}.

Differential variability was first examined, comparing the six major cancers to their corresponding normal tissue samples. Low variability CpG sites, defined to be those with standard deviation of beta values $<$0.05 in the combined cancer and matched normal samples, were considered to be noninformative and filtered out before analyses. A regression based approach is undertaken to assess homogeneity of variances of methylation M values between cancer and normal samples, essentially the classical Levene test for differential variability \citep{Brown1974,Phipson2014} and adjusting for age (and gender when applicable). Figure \ref{volcano_plot_six_cancers} shows the volcano plots for genome-wide DVC test results in the six cancers, where the x axis is the difference of standard deviation of methylation beta values. At the family-wise error rate (FWER) 0.05, there are 21220 (PCa), 54734 (CRC), 76455 (LUSC), 31817 (LUAD), 120073 (BRCA), and 132797 (HCC) significant DVCs for the six cancers respectively. Nearly all DVCs show increased variability in cancer samples ($>$99\% in BRCA, CRC, HCC, LUSC and LUAD, $>$95\% in PCa). Compared to the number of significant DMCs (Table \ref{all_DMresults}), liver cancer has more DVCs and breast cancer has a similar number of DVCs (Table \ref{all_DMresults}), probably because of several distinctive subtypes in these two cancers.

\begin{sidewaystable}
\caption{Results of DMC, DVC and two constrained tests for the six cancers in TCGA.}
\label{all_DMresults}
\centering
      {\begin{tabular}{@{}llcccccc@{}}
        \hline
            &   & PCa & BRCA   & CRC  & HCC & LUSC   &  LUAD \\ \hline
Data    & \# tumor samples & 498& 782 & 296 & 377& 370 & 548 \\
        & \# normal samples & 50&  96 &  38 & 50 & 42 & 32 \\
        & \# probes pass filtering & 216605 &  242402 & 254297 & 257721 & 262820 & 241668\\
DMC     & \# significant CpGs &  94411 & 120345 & 70114 & 94284 & 110930 & 61590 \\
        & \# significant hypermethylated CpGs & 58032 & 63219 & 35092 & 10757 &32971 & 29601 \\
        & \# significant hypermethylated CpGs  & &  & & & & \\
        &  \hspace{10pt} with mean beta in normals$<$0.1 & 8755 & 16210 & 8138 & 3064 & 10058 & 5228\\
DVC     & \# significant CpGs & 21220 & 120073 & 54734 & 132797 & 76455 & 31817\\
        & \# significant hypervariable CpGs & 20446 & 119684 & 54660 & 132685 & 76438 & 31813\\
        & \# significant DVC and significant DMC &  16014 & 76270 & 24023 & 69235 & 45066 & 15336\\
           & Proportion of DVC not significant for DMC &  25\% & 36\% & 56\% & 48\% & 41\% & 52\%\\
        & OR(DMC, DVC) & 4.6 & 3.1 & 2.6 & 4.3 & 2.6 & 3.3 \\
Constrained H$_{1b}$& \# significant CpGs & 108093 & 124072 & 107598  & 165292  & 175967  & 179143 \\
Constrained H$_{1c}$& \# significant CpGs &  70052 & 149055 & 94534 & 141082 & 100032 & 73440 \\
        & \# significant CpGs with low beta in normal& 10773 & 18041 & 10378 & 9316 &12846 &  7446\\
     \hline
      \end{tabular}}
\end{sidewaystable}



\begin{figure}[!ht]

  \vspace{15pt}
\centering
  \includegraphics[width=.75\textwidth]{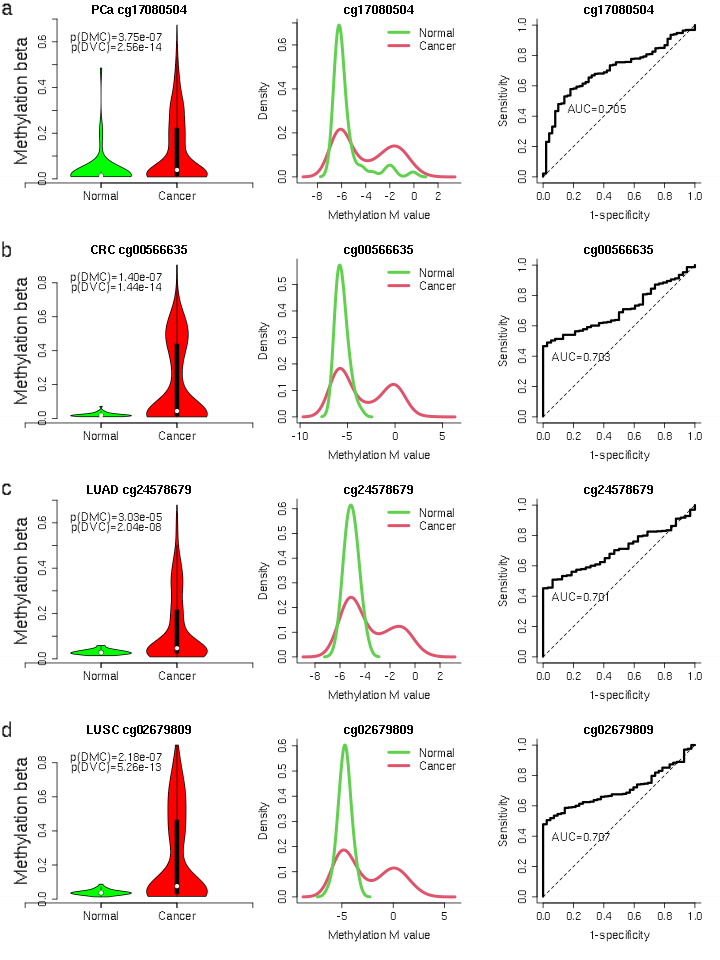}
        \caption{Four examples of significant DVC but not significant DMC at FWER $<$ 0.05. Violin plots, density plots and ROC curves were shown for each chosen CpG. (a) cg17080504 for prostate cancer. (b) cg00566635 for colorectal cancer. (c) cg24578679 for lung adenocarcinoma. (d) cg02679809 for lung squamous cell carcinoma. }
\label{four_examples}
      \end{figure}

A large portion of significant DVCs were not detected as significant DMCs, varying from  25\% to 53\% across 6 cancers (Table \ref{all_DMresults}), which suggests that testing for DVCs can increase the chance of detecting cancer aberrant methylation beyond DMCs. Figure \ref{four_examples} shows four examples of significant DVCs with low methylation in normal samples, but not detected as significant DMCs. In all CpG examples, violin plots for beta values and the density plots for M values show that there is a hypermethylated subgroup separating from the rest of samples that have similar methylation as the normal samples. These CpGs indexing heterogenous cancer subgroups may be better detected by testing for differential variability, since the p-values for testing DVCs are typically much smaller than the p-values for DMC. Interestingly, every one of these CpGs can discriminate cancer samples from normal samples with AUC$>$0.7, achieving a sensitivity $>$0.4 at nearly 100\% specificity. These DVCs have potential to delineate the heterogeneity in cancer methylome, and to be candidate cancer early detection markers when combined in a multi-marker panel.

We used the two constrained joint tests for differential mean and increased variable (hypervariable) as expressed in H$_{1b}$ and H$_{1c}$ for the six TCGA cancer-normal comparisons. For testing H$_{1b}$, a substantial higher amount of CpGs show FWER $<$0.05 than the numbers for DMC, e.g., nearly three times as much as DMCs for LUAD, 50\% more for LUSC.  In biomarker discovery studies, the interest is often more on the CpGs that are not methylated in normal samples (mean beta $<$0.1) to protect specificity, but hypermethylated in cancer samples. Across 6 comparisons, the constrained joint test for  H$_{1c}$ identified substantially more significant hypermethylated CpGs than the standard DMC test, generally 20-40\% more and as many as 3-fold drastic increase for HCC (Table \ref{all_DMresults}). This is perhaps not surprising given we observed the widespread increase of variances in cancer CpGs in Figure \ref{volcano_plot_six_cancers}.

\section{Discussion}
\label{sec4}

Testing for differential methylation between two sample groups has been the analytical workhorse of methylation studies. The methodological contribution of this work is integrating differences of means and increased variances into joint constrained hypothesis tests, motivated by the observation that DVCs are predominantly hypervariable in all six cancers in TCGA (Figure \ref{volcano_plot_six_cancers}). This strategy improves the power to identify cancer-specific CpGs in a genome-wide interrogation. As illustrated by the TCGA data example, exploiting increased variances increases the yield of candidate CpG markers.

When studying DVCs in the six common cancers from TCGA, one of the most interesting observations is that these DVCs often present heterogenous subgroups in cancer patients (Figure \ref{four_examples}), that may not been detected as DMCs with adequate significance due to multiple-testing adjustment. This reiterates the importance of accounting for cancer heterogeneity and subgroups when studying molecular markers. Similar objectives have been considered by using pAUC for detecting the presence of cancer subgroup for gene expression \citep{Pepe2003}. In simulations, we showed that accurate variance estimate of pAUC typically requires a large sample size, that testing pAUC in a high-dimensional setting may have an inflated type I error rate. In contrast, the proposed joint test with constrained hypothesis has better small-sample behaviors and consistently deliver the best power in diverse simulation scenarios.


Another potential utility of the joint test of differential mean and increased variance for DNA methylation is to use the derived CpGs for clustering analysis to define cancer subtypes. This objective has been universal for large-scale, genome-wide cancer methylation analyses, e.g. TCGA analyses. The standard starting point for clustering analysis is to extract the most variable top 5000 CpGs and feed them into a clustering algorithm. The constrained tests based on the cancer-normal comparison may yield a more informative set of cancer-specific markers, because they may capture the subgroups as we showed in Figure \ref{four_examples}.


\section*{Acknowledgments}

The authors thank Janet Stanford and Ming Yu for their helpful discussions that led to an improved version of the paper. This work is supported by National Institute of Health R01 CA222833 and computing support from National Institute of Health grant S10OD028685.

\subsection*{Conflict of interest}

The authors declare no potential conflict of interests.

\section*{Software}
\label{sec5}

The code for implementing the joint constrained test for increased mean and increased variance is included in the \texttt{DMVC} function in {\bf R} package \texttt{DMtest}, which is publicly accessible in CRAN ( {\tt https://cran.r-project.org/web/packages/DMtest/index.html}).

\bibliographystyle{rss}
\bibliography{DMtest}



\end{document}